\documentstyle[aps,epsfig,prl,multicol]{revtex}

\begin{document}
\vspace{0.0cm}
\draft

\title{Unconventional decay law for excited states in closed
many-body systems}

\author{V.V. Flambaum $^1$ \thanks{email address:
flambaum@newt.phys.unsw.edu.au} and F.M. Izrailev $^2$}

\address{$^1$ School of Physics, University of New South Wales,
Sydney 2052, Australia\\ $^2$Instituto de F\'{\i}sica, Universidad
Aut\'onoma de Puebla, Apartado Postal J-48, Puebla 72570,
M\'exico}

\date{\today}
\maketitle

\begin{abstract}

We study the time evolution of an initially excited many-body
state in a finite system of interacting Fermi-particles in the
situation when the interaction gives rise to the ``chaotic''
structure of compound states. This situation is generic for highly
excited many-particle states in quantum systems, such as heavy
nuclei, complex atoms, quantum dots, spin systems, and
 quantum computers. For a strong interaction the leading
term for the return probability $W(t)$ has the form $W(t)\simeq
\exp (-\Delta _E^2t^2)$ with $\Delta _E^2$ as the variance of the
strength function. The conventional exponential linear dependence
$W(t)=C\exp (-\Gamma t)$ formally arises for a very large time.
However, the prefactor $C$ turns out to be exponentially large,
thus resulting in a strong difference from the conventional
estimate for $W(t)$.

\end{abstract}

\pacs{PACS numbers:  03.67.Lx, 05.45.Mt, 24.10.Cn}

\begin{multicols}{2}

It is known that highly excited states can be treated as
``chaotic" ones in many-body systems, such as complex atoms
\cite{Ce}, multicharged ions \cite {ions}, nuclei \cite{nuclei}
and spin systems \cite{Nobel,spins}. This happens due to a very
high density of many-particle states, which strongly increases
with an increase of energy. For example, in the case of $n$
Fermi-particles occupying the finite number $m$ of ``orbitals"
(single-particle states), the total number $N$ of many-body states
grows exponentially fast with an increase of the number of
particles, $N =m!/n!(m-n)!\sim \exp(c_0 n)$. Correspondingly, the
density $\rho_f$ of those many-body states which are directly
coupled by a two-body interaction, also grows very fast
\cite{alt}. Therefore, even a relatively weak interaction between
the particles can lead to a strong mixing between unperturbed
many-body states (``basis states"). As a result, an exact
(perturbed) eigenstate is represented by a chaotic superposition
of a large number of components of basis states.

The number of principal basis components in such {\it chaotic}
eigenstates can be estimated as $N_{pc}\sim \Gamma /D$ where
$\Gamma $ is the {\it spreading width} of a typical component,
that can be estimated using the Fermi golden rule, and $D^{-1}(E)$
is the total density of many-body states. In the case of a quantum
computer, the interval between multiqubit energy levels $D\propto
1/N$ is extremely small, and practically it is impossible to
resolve these levels. Moreover, both the temperature and finite
time of computer operations lead to an energy uncertainty $\delta
E\gg D$. A similar situation occurs for an electron which enters a
many-electron quantum dot. In these cases the analysis of
stationary chaotic eigenstates is not an adequate to real physical
problems and one needs to consider the time evolution of wave
functions. In this Letter we extend the quantum chaos approach to
the problem of time evolution of an initially excited basis state.

Exact many-body eigenstates $\left| k\right\rangle \,$ of the
Hamiltonian $H=H_0+V$ of interacting Fermi-particles can be
expressed in terms of simple {\it shell-model basis states}
$\left| f\right\rangle $ of $ H_0$,
\begin{equation}
\label{slat}\left| k\right\rangle =\sum\limits_f C_f^{(k)}\left|
f\right\rangle \,;\,\,\,\,\,\,\,\,\left| f\right\rangle
=a_{f_1}^{+}...a_{f_n}^{+}\left| 0\right\rangle.
\end{equation}
Here $\left|0\right\rangle$ is the ground state, $a_s^{+}$ is the
creation operator and $C_f^{(k)}$ are components of an exact
eigenstate in the unperturbed basis.

In application to quantum computer models the Hamiltonian $H_0$
describes a number of non-interacting {\it qubits} (two-level
systems), and $V$ stands for the interqubit interaction needed for
a quantum computation ( we assume  time-independent $V$). In this
case the basis state $|f>$ is a product of single qubit states,
$a_s^{+}$ is the spin-raising operator (if the ground state
$\left|0\right\rangle$ corresponds to spins ``down"), and chaotic
eigenstates $\left| k\right\rangle $ are formed by the residual
interaction $V$.

Below we consider the time evolution of the system, assuming that
initially ($t=0$) the system is in a specific basis state $\left|
i\right\rangle $ (in the state with certain spins ``up'' for a
quantum computer). This state can be expressed as a sum over exact
eigenstates,
\begin{equation}
\label{in}\left| i\right\rangle =\sum\limits_kC_i^{(k)}\left|
k\right\rangle,
\end{equation}
therefore, the time-dependent wave function reads as
\begin{equation}
\label{psit}\Psi (t) =\sum\limits_{k,f}C_i^{(k)}C_f^{(k)}\left|
f\right\rangle \exp(-i E^{(k)}t).
\end{equation}
Here $E^{(k)}$ are the eigenvalues corresponding to the
eigenstates $|k>$. The sum is taken over the eigenstates $|k>$ and
basis states $|f>$ (in what follows, we put $\hbar=1$).

The probability $W_i=|A_i|^2 =|\left\langle
i|\Psi(t)\right\rangle|^2$ to find the system in the state $|i>$
is determined by the amplitude
\begin{eqnarray}
\label{ampli}
A_i= \left\langle i|\exp(-iHt)|i\right\rangle=
\sum\limits_k|C_i^{(k)}|^2\exp(-i E^{(k)}t) \simeq
\nonumber \\
\int P_i(E)\exp(-i Et)dE.
\end{eqnarray}
Here we replaced the summation over a large number of the
eigenstates by the integration over their energies $E \equiv
E^{(k)}$, and introduced the {\it strength function} (SF) $P_i(E)$
which is also known in the literature as the {\it local spectral
density of states},
\begin{equation}
\label{strength}P_i(E)\equiv \overline{|C_i^{(k)}|^2}\rho (E).
\end{equation}
Here $\rho (E)$ is the density of states of the total Hamiltonian
$H$, and the average is performed over a number of states with
close energies.

In chaotic systems the strength function $P_i(E)$ is known to have
the Breit-Wigner form for a relatively weak interaction, and is
close to the Gaussian for a strong interaction \cite{Ce,nuclei}.
Recently, the following approximate general expression  has been
analytically found \cite{FI00},
\begin{equation}
\label{FfBW} P_i(E) = \frac{1}{2\pi}\frac{\Gamma_i(E)}
{(E_i+\delta_i -E)^2+\Gamma_i(E)^2/4} ,
\end{equation}
\begin{equation}
\label{GammaH}\Gamma _i(E)\simeq 2\pi \overline{\left|
V_{if}\right| ^2}\rho_f(E).
\end{equation}
which is derived by making use of the approach described in
Ref.\cite{BM69}. Here $\Gamma_i(E)$ is some function of the total
energy, $\delta_i$ is the correction to the unperturbed energy
level $E_i$ due to the residual interaction $V_{if}$, and
$\rho_f(E)$ is the density of the basis states
$\left|f\right\rangle$ directly connected with a given state
$\left| i\right\rangle$ by the matrix element $V_{if}$. The above
result has been derived for the so-called Two-Body Random
Interaction (TBRI) model \cite{old} which describes $n$
interacting Fermi-particles distributed over $m$ orbitals, with an
assumption that two-body matrix elements are completely random.

It is shown \cite{FI00} that for a large number of particles the
function $\Gamma _i(E)$ has the Gaussian form with the variance
which depends on the model parameters. In the case of a relatively
small (but non-perturbative) interaction (when $\Gamma _i\equiv
2\pi \rho _fV_0^2\ll 2\Delta _E$ with $ V_0^2=\langle
V_{if}^2\rangle $), the function $\Gamma (E)$ is very broad (i.e.
it does not change significantly within the energy intervals
$\sim$ $ \Gamma$ and $\Delta_E$) and can be treated as constant,
$\Gamma (E)\simeq \Gamma _0$. In the other limit case of a strong
interaction, $\Gamma _0\ge \Delta _E$, the dependence $\Gamma (E)$
in (\ref{FfBW}) is the leading one.

The knowledge of the strength function allows one to describe the
dynamics of wave packets in the energy space. It is easy to find
the evolution of $ W_i(t)$ on a small time scale. Let us subtract
the energy $E_i\equiv H_{ii}$ of the initial state in the exponent
and make a second order expansion in $ E-E_i$ in Eq.(\ref{ampli}).
This gives the following result,
\begin{equation}
\label{At2}A_i=\exp (-iE_it)\left( 1-\Delta _E^2t^2/2\right)
\end{equation}
and
\begin{equation}
\label{Wt2}W_i(t)=1-\Delta _E^2t^2.
\end{equation}
Here the width $\Delta _E$ of the SF is determined through the
second moment, $\Delta _E^2=\sum\limits_{f\neq i}V_{if}^2$, which
for the TBRI model is \cite{FI97},
\begin{equation}
\label{SFwidth}\Delta _E^2=\frac 14V_0^2n(n-1)(m-n)(m-n+3),
\end{equation}
with $V_0^2$ standing for the variance of the off-diagonal
elements ($\Delta _E^2$ for a quantum computer model can be found
in Ref.\cite{F2000}).

Note that for a strong residual interaction, $\Gamma_0 \ge
\Delta_E$, the time dependence (\ref{Wt2}) is also correct for a
longer time \cite{F2000}. Indeed, both the strength function and
density of states in this limit are described by the Gaussian
functions with the variance $\sigma^2=\Delta_E^2$ (see details in
\cite{old} - \cite{Kota2000}),
\begin{equation}
\label{PGauss} P_i(E) = \frac{1}{\sqrt{2\pi \sigma^2}}
\exp\left[-\frac{(E-E_c)^2}{2\sigma^2}\right],
\end{equation}
\begin{equation}
\label{rhoGauss} \rho(E) = \frac{N}{\sqrt{2\pi \sigma^2}}
\exp\left(-\frac{E^2}{2\sigma^2}\right).
\end{equation}
Therefore, at the center of the energy spectrum, $E_c=0$,
Eq.(\ref{ampli}) results in the Gaussian time dependence for
$A_i(t)$ and $W_i(t)$,
\begin{equation}
\label{Alarge} A_i= \exp\left(-\Delta_E^2 t^2/2\right),
\end{equation}
\begin{equation}
\label{Wlarge} W_i(t)\simeq\exp\left(- \Delta_E^2 t^2\right).
\end{equation}

Now let us consider large times. In this limit the result can be
obtained by evaluation of the integral in Eq.(\ref{ampli}) in the
complex $E$-plane. Specifically, one should close the contour of
integration in the bottom part of the complex plane ($Im\,E<0$),
in order to provide a vanishing contribution at infinity. Then,
the large time limit is given by the pole of the strength function
(\ref{FfBW}), closest to the real $E$-axis. If $\Gamma_i$ and
$\delta_i$ in Eq.(\ref{strength}) do not depend on $E$, the
integration gives the conventional exponential decay
$W_i=\exp(-\Gamma t)$ \cite{BM69}. However, the energy dependence
of $\Gamma$ is necessary to provide the finite second moment
$\Delta_E^2$ of the strength function. If $\Gamma <\Delta_E$, the
closest pole is given by $\tilde\Gamma= - 2\, Im\,E_p$, where
$E_p$ is the solution of the equation $E_p=E_i +\delta_i(E_p) -
i\Gamma(E_p)/2$ with a minimal imaginary part. If $\Gamma
\ll\Delta_E$, then we have $\tilde\Gamma\approx\Gamma$. As a
result, we obtain an exponential dependence for large $t$,
\begin{equation}
\label{Wtinf} W_i(t)=C \exp(- \tilde\Gamma t),
\end{equation}
with some constant $C$.

In Ref.\cite{F2000} simple extrapolation formula for $W_i(t)$ has
been suggested, that interpolates (for $\Gamma <\Delta _E$)
between the small (\ref{Wt2}) and large (\ref{Wtinf}) time
dependencies,
\begin{equation}
\label{Wint}W_i(t)=\exp \left( \frac{\Gamma ^2}{2\Delta _E^2}-\sqrt{\frac{
\Gamma ^4}{4\Delta _E^4}+\Gamma ^2t^2}\right) .
\end{equation}
The transition between the Gaussian regime and simple exponential
decay occurs near the time $t_c\sim \Gamma /\Delta _E^2$ where
$\Delta _E^2t_c^2\sim \Gamma t_c$. This gives an estimate,
$C\sim\exp \left( \frac{\Gamma ^2}{2\Delta _E^2}\right) $, for the
constant $C$. Indeed, for $t<t_c$ the quadratic exponential decay,
$\exp \left( -\Delta _E^2t^2\right) $, is slower than the linear
one, $\exp (-\Gamma t)$. The matching of these two dependencies
would naturally require the above expression for $C$. Thus, the
constant $C$ can be large if $\Gamma >\Delta _E$. The transition
from one regime of the time dependence of $W_i(t)$ to another is
schematically shown in Fig.1.

\begin{figure}[htb]
\vspace{-1.0cm}
\begin{center}
\hspace{-1.5cm}
\epsfig{file=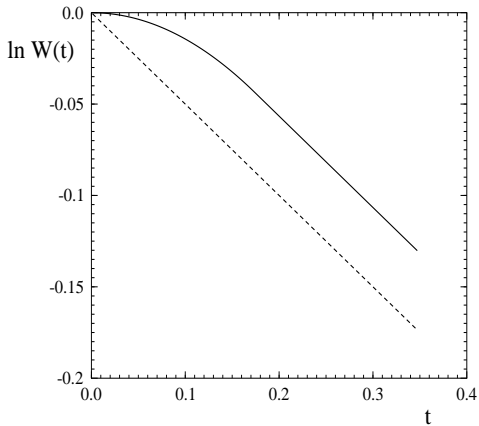,width=3.0in,height=2.3in,angle=-90}
\vspace{-0.8cm}
\caption{
Schematic time dependence $W(t)$ for $\tilde\Gamma=0.5$,
$\Delta_E=1.2$; the dependence $W(t)=\exp(-\Delta_E^2 t^2)$
changes into $ W(t)=\exp(-\Gamma t)$ at the point
$t_c=\Gamma_p/\Delta_E^2\approx 0.17$. Simple exponential
dependence is shown by the dotted line.}
\end{center}
\end{figure}

This qualitative result is supported by a more detailed
consideration. In many-body systems the dependence $\Gamma (E)$
can be approximated by the Gaussian function since the density of
final states $\rho_f(E_f)$ in Eq.(\ref{GammaH}) is typically close
to the Gaussian \cite{FI00}. This provides us with an approximate
formula for the strength function (tested by the numerical
calculations \cite{casa}),
\begin{equation}
\label{approx} P(E)=B\frac{ \exp\left[-\frac{(E-E_0)^2}
{2 \sigma^2}\right]}{(E-E_0)^2+\Gamma^2/4}.
\end{equation}
It can be used, in conjunction with Eq.(\ref{ampli}), to study the
time dependence $W_i(t)$. Strictly speaking, this formula is valid
near the center of the energy spectrum, otherwise one should take
into account distortion effects near the edges of the energy
spectrum.

Due to the normalization conditions, $\int P(E) dE =1 $ and $\int
E^2 P(E) dE = \Delta_E^2$, we have the following relations
\cite{casa},
\begin{equation}
\label{18}\frac 1B=2\left[ 1-\Phi \left( \frac \Gamma {\sigma\sqrt{8}
}\right)\right] \frac \pi \Gamma \exp \left(\frac{\Gamma
^2}{8\sigma ^2}
\right),
\end{equation}
and
\begin{equation}
\label{19}\Delta_E^2=B\left\{ \sigma\sqrt{2\pi }
-\frac{\pi \Gamma }2\exp
\left( \frac{\Gamma ^2}{8\sigma ^2}\right)
\left[ 1-\Phi \left(\frac \Gamma
{\sigma\sqrt{8}}\right)\right] \right\}
\end{equation}
where $\Phi (z)$ is the error function.

The return probability $W(t)$ corresponding to the strength
function (\ref{approx}) is defined by the integral,
\begin{equation}
\label{20}A(t)=B\int\limits_{-\infty }^\infty dE\frac{\exp
\left( -\frac{E^2}{2\sigma ^2}-iEt\right) }{\left( E-E_0\right) ^2
+\frac{\Gamma ^2}4}
\end{equation}
The time dependencies of $A(t)$ and $W(t)$ for small time are
given by Eqs.(\ref{Alarge}-\ref{Wlarge}). If $\Gamma \ll \sigma $,
the region of the applicability of these equations is very narrow.
Indeed, in this case $\Delta_E^2\approx \frac{\sigma \Gamma
}{\sqrt{2\pi }}\,$ and the condition $ t\ll t_c\ll \frac \Gamma
{\Delta_E^2}\sim \frac 1\sigma $ results in the relation
$\Delta_E^2t^2\ll 1$ . The absolute value of the
 amplitude $A(t)$ in this case is given
by the series in the parameter $(\sigma t)^2=(t/t_c)^2$ ,
\begin{equation}
\label{21}|A(t)|=1-\frac 12\frac{\Gamma \sigma }{\sqrt{2\pi }}t^2
+\frac 1{24}\frac{\Gamma \sigma ^3}{\sqrt{2\pi }}t^4\,+\,...
\end{equation}

For large time, $t\gg t_c=1/\sigma $, the calculation of the
integral in Eq.(20) gives $$ W(t)\approx\exp\left( \frac 1\pi
\frac{\Gamma ^2}{\Delta_E^2}-\Gamma t\right) $$ for the return
probability. Here the correction $\frac 1\pi \frac{\Gamma
^2}{\Delta_E^2}\approx \frac{2\Gamma }{\sigma \sqrt{2\pi }}$ is
small. Indeed, the strength function in this case is close to the
Lorentzian which gives a simple dependence $W(t)=\exp (-\Gamma t)$
for the decay.

Another limit case of a strong interaction, $\Gamma \gg \Delta_E$
(or, the same, $\Gamma\gg\sigma $), is more delicate. In this case
the strength function is close to the Gaussian with $\Delta_E
\approx \sigma $ and $t_c$ is large, $t_c\sim \frac \Gamma {\sigma
^2} \gg \frac 1\sigma $. The leading dependence of $W(t)$ in this
case is the Gaussian, $W(t) \simeq \exp (-\Delta_E^2t^2)$ . Only
for a long time $t\gg \frac \Gamma {\sigma ^2}$ it becomes the
simple exponential function,
\begin{equation}
\label{Wfinal}W(t)\approx \frac{\pi ^2\Gamma ^2}
{8\Delta_E^2}\exp \left(
\frac{1}{4} \frac{\Gamma ^2}{\Delta_E^2}-\Gamma t\right)
\end{equation}
It is important to stress that even for a large time the return
probability $W(t)$ has large correction factor $\exp \left(
\frac{1}{4}
\frac{\Gamma ^2}{ (\Delta E)^2}\right) $, in addition to the
standard decay law $\exp (-\Gamma t)$.

Due to a finite number of particles, there are specific important
features in the dynamics of wave packets, namely, the damped
oscillations and the break of the decay for $W_i(t)$
\cite{FI01}. The number of basis components $\left| f\right\rangle
$ within the energy shell $\left| E_0-E_f\right|\leq\min (\Gamma
,\sigma )\equiv\Delta$ is finite. Therefore, the decay stops if
$W_i\,$ is close to the equilibrium value defined as $W_\infty
\equiv\overline{W_i(t\rightarrow \infty )}\approx 3N_{pc}^{-1}$.
Here $N_{pc}$ is the number of principal components in an
eigenstate, $N_{pc}\sim \Delta /D$ , where $D=\rho^{-1}$ is the
mean energy interval between all many-body levels. Note that the
value of $W_\infty $ is still at least 3 times larger than
$W_f=N_{pc}^{-1}$ for any other component $f$ ,
$\overline{W_i(\infty )}\geq $ $\overline{W_{f}(\infty )}\,$ (see,
details in \cite{FI01}).

The equilibrium occurs because the average decay flux is equal to
the average return flux. However, the return flux leads to the
damped oscillations of $W_i(t)$ and to the oscillations of a
current number of the principal components $N_{pc}(t)$. These
oscillations arise because the decay flux ``reflects'' from the
edges of the energy shell, when all components within this shell
are populated. Period of these oscillations is about $ n_c/\Delta
$ where $\Delta $ is the inverse decay time, and $n_c$ is the
number of ``classes'' in the Hilbert space. This number can be
defined as the number of interaction steps in the perturbative
chain $H_{0\alpha _1}H_{\alpha _1\alpha _2}\,.\,.\,.\,H_{\alpha
_k\alpha _{n_c}}$, needed to populate all basis states within the
energy shell. For example, in the TBRI model with 6 particles and
$12$ orbitals, the number of steps is $n_c=3$ since each two-body
interaction $H_{ik}$ moves only two particles to new orbitals.

In conclusion, we have studied generic features of the return
probability $ W(t)$ for a system to be found in an initially
excited many-body state. Due to the two-body interaction between
Fermi-particles, the wave packet in the energy representation
spreads over all basis states within the energy shell. The
dependence $W(t)$ for small time is determined by a ballistic
spread of the packet and is given by the expression (\ref{Wt2}).
For large time, the decrease of $W(t)$ is determined by the form
of the strength function $P(E)$ . We have analyzed the behavior of
$W(t)$ by making use of the analytical expression for $P(E)$,
which is obtained for any strength of random two-body interaction
between finite number of interacting Fermi-particles.

We have shown that for the Breit-Wigner form of $P(E)$ (relatively
weak interaction) the decay of $W(t)$ on a large time scale has
the conventional exponential dependence, $W(t) \simeq \exp
(-\Gamma t)$. On the other hand, for the Gaussian form of $P(E)$
(strong interaction) the time dependence $W(t)$ turns out to be of
very specific. Namely, the leading term gives the quadratic
exponential dependence, $W(t)\sim \exp (-\Delta _E^2t^2)$, and
only for a very large time the conventional exponential linear
dependence formally recovers. However, in this case an additional
prefactor $C$ appears before the exponent, which turns out to be
exponentially large, thus resulting in a strong modification of
the standard exponential estimate for $W(t)$.

This work was supported by the Australian Research Council. One of
us (F.M.I.) gratefully acknowledges the support by CONACyT
(Mexico) Grant No. 34668-E. The authors are grateful to M.Yu.
Kuchiev for valuable discussions.

\end{multicols}

\begin{thebibliography}{99}
\bibitem{Ce}  V. V. Flambaum, A. A. Gribakina, G. F. Gribakin, and M. G.
Kozlov, Phys. Rev. {\bf A 50}, 267 (1994).

\bibitem{ions}  G.F. Gribakin, A.A. Gribakina, V.V. Flambaum. Aust. J.Phys.
{\bf 52}, 443 (1999).

\bibitem{nuclei}  M.Horoi, V.Zelevinsky and B.A.Brown, Phys. Rev. Lett. {\bf %
74}, 5194 (1995); V.Zelevinsky, M.Horoi and B.A.Brown, Phys. Lett. {\bf B}
{\bf 350}, 141 (1995); N.Frazier, B.A.Brown and V.Zelevinsky, Phys. Rev.
{\bf C} {\bf 54}, 1665 (1996); V.Zelevinsky, B.A.Brown, M. Horoi and
N.Frazier, Phys. Rep., {\bf 276} , 85 (1996).

\bibitem{Nobel}  V. V. Flambaum, Proc. 85th Nobel Symposium,
Phys. Scr. {\bf 46}, 198 (1993).

\bibitem{spins}  B.~Georgeot and D.~L.~Shepelyansky, Phys. Rev. Lett.
{\bf 81},5129 (1998).

\bibitem{alt}  S. Aberg. Phys. Rev. Lett. {\bf 64}, 3119 (1990).
D.L.Shepelyansky and O.P.Sushkov, Europhys. Lett. {\bf 37}, 121
(1997); B.L.Altshuler, Y.Gefen, A.Kamenev and L.S.Levitov, Phys.
Rev. Lett., {\bf 78}, 2803 (1997); A.D.Mirlin and Y.V.Fyodorov,
Phys. Rev. {\bf B 56}, 13393 (1997); D.Weinmann, J.-L. Pichard and
Y.Imry, J.Phys. I France, {\bf 7}, 1559 (1997); P.Jacquod and
D.L.Shepelyansky, Phys. Rev. Lett. {\bf 79} , 1837 (1997);
V.V.Flambaum and G.F.Gribakin, Phys. Rev. {\bf C 50}, 3122 (1994);
P.G.Silvestrov, Phys. Rev. Lett., {\bf 79}, 3994 (1997); Phys.
Rev. E. {\bf 58}, 5629 (1998).

\bibitem{FI00}  V.V.Flambaum and F.M.Izrailev, Phys. Rev. E.,
{\bf 61}, 2539 (2000).

\bibitem{BM69}  A. Bohr and B. Mottelson, {\em Nuclear structure, Vol. 1}
(Benjamin, New York, 1969).

\bibitem{old}  J. B. French and S. S. M. Wong, Phys. Lett.
{\bf B} {\bf 35}, 5 (1970); O. Bohigas and J. Flores, Phys. Lett.
B {\bf 34}, 261 (1971).

\bibitem{FI97}  V.V.Flambaum and F.M.Izrailev, Phys. Rev. {\bf E 55},
R13 (1997); {\bf E 56}, 5144 (1997).

\bibitem{F2000}  V.V. Flambaum. Aust. J.Phys. {\bf 53}, N4, (2000).

\bibitem{brody}  T.A. Brody, J. Flores, J.B. French, P.A. Mello,
A. Pandey, and S.S.M. Wong, Rev. Mod. Phys. {\bf 53}, 385 (1981).

\bibitem{Kota2000}  V.K.B. Kota and R. Sahu, nucl-th/0006079.

\bibitem{casa}  G.Casati, V.V.Flambaum, and F.M.Izrailev,
to be published.

\bibitem{FI01}  V.V.Flambaum and F.M.Izrailev, to be published.
\end{thebibliography}
\end{document}